\documentclass[usenatbib]{mn2e} 
\usepackage{epsfig}

\newif\ifAMStwofonts
\AMStwofontstrue

\def\be{\begin{equation}}
\def\ee{\end{equation}}

\def\Msun{M_{\odot}}
\def\Zsun{Z_{\odot}}

\def\gtsima{$\; \buildrel > \over \sim \;$}
\def\ltsima{$\; \buildrel < \over \sim \;$}
\def\prosima{$\; \buildrel \propto \over \sim \;$}
\def\gsim{\lower.5ex\hbox{\gtsima}}
\def\lsim{\lower.5ex\hbox{\ltsima}}
\def\simgt{\lower.5ex\hbox{\gtsima}}
\def\simlt{\lower.5ex\hbox{\ltsima}}
\def\simpr{\lower.5ex\hbox{\prosima}}

\title[Ly$\alpha$ versus X-ray heating]{Ly$\alpha$ versus X-ray heating in the high-z IGM}

\author[Ciardi, Salvaterra \&  Di Matteo]{B. Ciardi$^1$, R. Salvaterra$^2$ \& T. Di Matteo$^3$\\
$^1$ Max-Planck-Institut f\"ur Astrophysik, Karl-Schwarzschild-Stra\ss e 1, 85748 Garching, Germany\\
$^2$ INAF-Osservatorio Astronomico di Brera, via E. Bianchi 46, 23807 Merate (LC), Italy\\
$^3$ McWilliams Center For Cosmology, Carnegie Mellon University, 5000 Forbes Avenue, Pittsburgh PA 15213, USA\\}

\date{May 09}    

\pagerange{\pageref{firstpage}--\pageref{lastpage}}
\pubyear{2009}
\begin{document}

\maketitle
\label{firstpage}

\begin{abstract}
In this paper we examine the effect of X-ray and Ly$\alpha$ photons on the
intergalactic medium temperature. We calculate the photon production from a
population of stars and micro-quasars in a set of cosmological hydrodynamic
simulations which self-consistently follow the dark matter dynamics, radiative
processes as well as star formation, black hole growth and associated feedback
processes.  We find that, {\it (i)} IGM heating is always dominated by X-rays
unless the Ly$\alpha$ photon contribution from stars in objects with mass
$M<10^8$~M$_\odot$ becomes significantly enhanced with respect to the X-ray
contribution from BHs in the same halo (which we do not directly model). {\it
(ii)} Without overproducing the unresolved X-ray background, the gas
temperature becomes larger than the CMB temperature, and thus an associated
21~cm signal should be expected in emission, at $z \simlt 11.5$. We discuss
how in such a scenario the transition redshift between a 21~cm signal in
absorption and in emission could be used to constraint BHs accretion and
associated feedback processes.
\end{abstract}


\section{Introduction}

While the investigation of the very high-$z$ universe, $z \sim 1000$, is
possible through the Cosmic Microwave Background (CMB) radiation and detection
of a handful of objects with redshift as high as $\sim 7$ has recently been
possible, there is a lack of observational data in the redshift interval $7
\simlt z \simlt 1000$. To detect radiation from this interval, telescopes with
exceptional sensitivity in the IR and radio bands are needed. {\tt
JWST}\footnote{http://ngst.gsfc.nasa.gov} (James Webb Space Telescope), for
example, with its nJy sensitivity in the $1-10$~$\mu$m infrared regime, is
ideally suited for probing optical-UV emission from sources at
$z>10$. Similarly, the planned generation of radio telescopes as {\tt
SKA}\footnote{http://www.skatelescope.org} (Square Kilometer Array), {\tt
LOFAR}\footnote{http://www.lofar.org} (LOw Frequency ARray), {\tt 21cmA} (21cm
Array) and {\tt MWA}\footnote{http://web.haystack.mit.edu/arrays/MWA/}
(Murchison Widefield Array) will open a new observational window on the high
redshift universe. In particular, the detection of the 21~cm line associated
with the hyperfine transition of the ground state of neutral hydrogen, holds
the promise to shed light on the reionization process and its sources.

This observational progress has been accompanied by a flourishing of
theoretical activity, aimed at providing predictions for the planned
generation of telescopes.  It has long been known (e.g. \citealt{Field_59})
that neutral hydrogen in the intergalactic medium (IGM) and gravitationally
collapsed systems may be directly detectable in emission or absorption against
the CMB at the frequency corresponding to the redshifted HI 21~cm line
(associated with the spin-flip transition from the triplet to the singlet
ground state). \citet{MadauMeiksinRees_97} first showed that 21~cm tomography
could provide a direct probe of the era of cosmological reionization and
reheating. In general, 21~cm spectral features will display angular structure
as well as structure in redshift space due to inhomogeneities in the gas
density field, hydrogen ionized fraction, and spin temperature. Several
different signatures have been investigated in the recent literature among
which fluctuations in the 21~cm line emission induced by the ``cosmic web''
\citep{Tozzi_etal_00}, by the neutral hydrogen surviving reionization
(e.g. \citealt{CiardiMadau_03,
FurlanettoSokasianHernquist_04,Mellema_etal_06}) and a global feature
(``reionization step'') in the continuum spectrum of the radio sky that may
mark the abrupt overlapping phase of individual intergalactic HII regions
\citep{Shaver_etal_99}.

A key feature for the observation of the line in emission is that the IGM
should be heated above the CMB temperature. While pre-heating can happen due
to, e.g., dark matter particles decay or annihilation
(e.g. \citealt{MapelliFerraraPierpaoli_06, Valdes_etal_07}), the main source
of heating are Ly$\alpha$ and X-ray photons. Although the impact of these
sources has been investigated by several authors (e.g.
\citealt{MadauMeiksinRees_97,ChenMiralda-Escude_04,Nusser_05a,KuhlenMadauMontgomery_06,
CiardiSalvaterra_07,PelupessyDiMatteoCiardi_07,RipamontiMapelliZaroubi_08}),
none of the above studies has approached the problem considering both sources
within a self-consistent model.  Here, we estimate the Ly$\alpha$ and X-ray
photon production using the simulations by \citet{PelupessyDiMatteoCiardi_07},
which were designed to follow the evolution of both quasar and stellar type
sources, including their associated feedback effects.
The above estimates are then used to calculate with a semi-analytic approach
the evolution of the temperature of an IGM at the mean density.

The paper is structured as follows. In Section~\ref{sec:simul} we describe the
main characteristics of the simulations by \citet{PelupessyDiMatteoCiardi_07},
while in Section~\ref{sec:phot_count} we give the prescription used to derive
the Ly$\alpha$ and X-ray photon count. In Section~\ref{sec:heating} we
estimate the IGM heating due to the emitted photons, in Section~\ref{sec:21cm}
the consequences for the observability of the 21~cm line from neutral hydrogen
and in Section~\ref{sec:summary} we give our conclusions.

\section{Simulations of halo collapse and black holes growth}
\label{sec:simul}

As mentioned in the Introduction, in this paper we make use of the simulations
described in \citet{PelupessyDiMatteoCiardi_07}. Here we briefly outline the
main characteristics and we refer the reader to the original paper for
details.

The simulations are designed to investigate the physical conditions for the
growth of intermediate mass seed black holes (BHs; possibly the remnants of a
first generation of massive stars), using the parallel cosmological
TreePM-Smooth Particle Hydrodynamics (SPH) code {\small
GADGET2}~\citep{Springel_05} in the standard $\Lambda$CDM
model\footnote{$\Omega_{\Lambda}=0.7$, $\Omega_m=0.3$, $\Omega_b=0.04$ and
$H_0=100 h$ km~s$^{-1}$~Mpc ($h=0.7$), where the symbols have the usual
meaning.}. The initial conditions for the simulation correspond to isolated
spherical overdensities ('top-hat') endowed with an appropriate Zeldovich
power spectrum (similar to \citealt{Bromm2004}).  During collapse of the
parent halo (all halos considered have $T_{\rm vir} > 10^4$ K) the seed holes
are incorporated through mergers into larger systems and accrete mass from the
surrounding gas. The interstellar medium (ISM), star formation and supernovae
feedback as well as black hole accretion and associated feedback are treated
self-consistently by means of sub-resolutions models. In particular, the
multiphase model for star forming gas has been developed by
\cite{SpringelHernquist_03}, while the prescription for accretion and feedback
from massive black holes is the one developed by
\cite{DiMatteoSpringelHernquist_05} and
\cite{SpringelDiMatteoHernquist_05}. Technically, black holes are represented
by collisionless particles that grow in mass by accreting gas from their
environments. The (unresolved) accretion onto the black hole is related to the
large scale (resolved) gas distribution using a Bondi-Hoyle-Lyttleton
parameterization \citep{Bondi_52, BondiHoyle_44}, and it is limited by the
Eddington rate. A fraction $\epsilon_{\rm f}$ of the radiative energy released
by the accreted material is assumed to couple thermally to nearby gas. In
addition, two black hole particles are assumed to merge if they come within
the spatial resolution of the simulation (i.e. within the local SPH smoothing
length) and their relative speed lies below the local sound speed. The seeding
procedure constists in selecting objects with a mass of $10^6$~M$_\odot$ and
place a seed BH of $M_{\rm seed} = 10^{3-4}$~M$_\odot$ in them if they do not
already contain a BH.

Different collapse scenarios have been investigated by choosing the total mass
$M$ of the host halos, the redshift of collapse $z_{\rm vir}$ and the spin
parameter $\lambda$. The halos considered have a mass of $M=10^{8}, 10^9,
10^{10}$ and $10^{11}$~M$_\odot$, with a virialization redshift of $z_{\rm
vir} = 16, 12, 10$ and $7.5$, respectively. A standard spin parameter of
$\lambda=0.03$ is adopted, but simulations with $\lambda = 0$ or
$\lambda=0.05$ have determined no significant difference in the results
\citep{PelupessyDiMatteoCiardi_07}. The simulations used here are the highest
resolution ones, having $N = 10^7$ particles (in dark matter and gas), with a
corresponding spatial resolution of a about hundred parsec. Two extreme values
for the BH feedback have been considered: $\epsilon_{\rm f}=0$ corresponding
to the case without any black hole feedback (this can be used for comparison)
and $\epsilon_{\rm f}=0.5$, the value that has been used to reproduce the
observed normalization of the local $M_{\rm BH}$-$\sigma$ relation in galaxy
merger simulations \citep{DiMatteoSpringelHernquist_05} as well as in full
cosmological hydrodynamical simulations where the same modeling was applied
\citep{DiMatteo2008}. All simulations here are run to a final redshift of
$z\sim 6$. 

Despite the analysis of a quite broad parameter space, the simulations do not
account for possible effects associated with gravitational recoil (for a
thourough discussion on the implications we refer the reader to the original
paper).

The simulations are used to derive the X-ray and Ly$\alpha$ photon production
as described in the following Section.

\section{Photon count}
\label{sec:phot_count}

In this Section we discuss our estimate of the X-ray and Ly$\alpha$ photon
production and associated background. Our reference run has $M_{\rm seed} =
10^4$~M$_\odot$, $\lambda=0.03$, $\epsilon_{\rm f}=0.5$.

The bolometric luminosity associated to a BH accreting at a rate $\dot{M}_{\rm
BH}(z,M_h)$ hosted in a halo of mass $M_h$ at redshift $z$ is
$L(z,M_h)=\epsilon c^2 \dot{M}_{\rm BH}(z,M_h)$, where $\epsilon=0.1$ is the
radiative efficiency for accretion (for a Shakura-Sunyaev, and non-spinning
black hole) and $c$ is the speed of light. The comoving specific emissivity at
redshift $z$ is then:

\begin{equation}
j_{\rm BH}(\nu,z)=\int_{M_{h,min}}^{M_{h,max}} \frac{l_{\rm BH}(\nu,z)}{\int l_{\rm BH}
(\nu^\prime,z)d\nu^\prime}  L(z,M_h) n_{\rm PS}(z,M_h) dM_h,
\label{eq:jBH}
\end{equation}

\noindent
where $n_{\rm PS}(z,M_h)$ is the weight for the halo of mass $M_h$ at redshift
$z$ as computed by the Press-Schechter formalism\footnote{Before the redshift
of virialization of the halo $z_{vir}(M_h)$, the weight has been computed at
$z_{vir}$.}, and $M_{h,min}$ and $M_{h,max}$ are the minimum and maximum
masses for our simulated halo mass, respectively.
$M_{h,min}$ is set by the physics appropriately captured in the simulations,
which do not follow the chemistry of molecular hydrogen.
$l(\nu,z)$ is the average spectrum energy distribution for AGNs as computed by
\cite{SazonovOstrikerSunyaev_04}.

The total cosmic star formation rate density at redshift $z$ is: 

\begin{equation}
\dot{\rho}_{\star}(z)=\int_{M_{h,min}}^{M_{h,max}} \dot{M}_{\star}(z,M_h) n_{\rm PS}(z,M_h),
\label{eq:sfr}
\end{equation}

\noindent
where $\dot{M}_{\star}(z,M_h)$ is the star formation rate for the halo of mass
$M_h$ at redshift $z$.

For stars, the comoving specific emissivity is computed by:

\begin{equation}
j_\star(\nu,z)= \int_z^{\infty} dz^\prime l_\star(\nu,t_{z,z^\prime}) \dot{\rho}_{\star}(z),
\end{equation}

\noindent
where $l_\star( \nu,t_{z,z^\prime})$ is the template specific luminosity for a
stellar population of age $t_{z,z^\prime}$ (time elapsed between redshift
$z^\prime$ and $z$) as computed by \cite{BruzualCharlot_03} for Pop~II stars
with $Z=0.2\;\Zsun$ and a Salpeter IMF.  In \citet{CiardiSalvaterra_07} it was
shown that, although the ionizing photon production is much reduced (by a
factor of $\sim 4$) compared to metal-free stars with the same IMF, the
Ly$\alpha$ photon emission is similar. For this reason we limit our discussion
to metal enriched stars.

The background intensity $J(\nu_0,z_0)$ seen at a frequency $\nu_0$ by an
observer at redshift $z_0$ is then given by:

\begin{equation}\label{eq:back}
J(\nu_{0},z_0)= \frac{(1+z_0)^3}{4\pi}\int^{\infty}_{z_{0}}
j(\nu,z)e^{-\tau(\nu_0,z_0,z)}\frac{dl}{dz}dz,
\end{equation}

\noindent
where $\nu=\nu_0(1+z)/(1+z_0)$, $dl/dz$ is the proper line element, and
  $j(\nu,z)=j_\star(\nu,z)+j_{\rm BH}(\nu,z)$ is the total emissivity.
  $\tau(\nu_0,z_0,z)$ is the optical depth of the medium.

\begin{figure}
\centering
\includegraphics[width=0.5\textwidth]{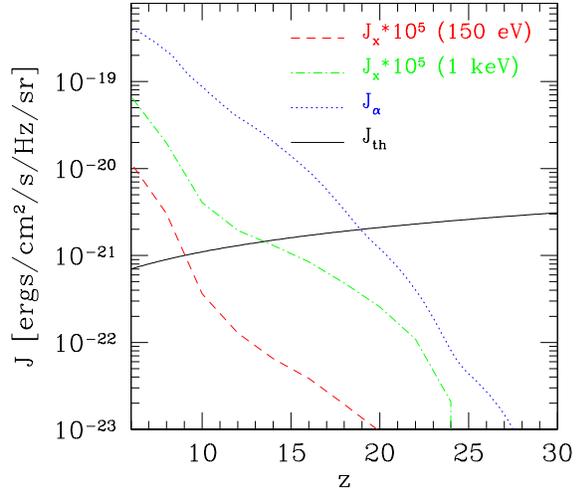}
\caption{Evolution of the Ly$\alpha$ (dotted line), X-ray at 150~eV (dashed)
and 1~keV (dashed-dotted) background radiation.  The solid line is the minimum
intensity required for the Ly$\alpha$ background to be effective in decoupling
the spin temperature from $T_{\rm CMB}$.  }
\label{fig:back}
\end{figure}

For Ly$\alpha$ photons, $\tau$ as been computed as in \citet[][see Sect.~2.2
for a full description of the IGM modeling]{SalvaterraFerrara_03}.
 The evolution of the Ly$\alpha$ background is shown in
Figure~\ref{fig:back}. We note that its intensity is 
much smaller than what found in
\citet{CiardiSalvaterra_07}. This effect can be attributed to the larger
$M_{h,min}$ employed here, as set by the smallest simulated halos.

In the X-rays,

\begin{equation}
\tau(\nu,z_0,z)=\int_{z_0}^z  \frac{dl}{dz} \sigma\left(\nu\frac{1+z^\prime}{1+z_0}\right) n_B(z^\prime),
\end{equation}

\noindent
where $\sigma(\nu)$ is the photon-ionization cross section per baryon of a
cosmological mixture of H and He (\citealt{ZdziarskiSvensson_89}; see also
\citealt{RipamontiMapelliZaroubi_08} for a more thorough discussion) and
$n_B(z)=n_B(0)(1+z)^3$ is the cosmological baryon number density at redshift
$z$ ($n_B(0)\simeq 2.5\times 10^{-7}$ cm$^{-3}$; Spergel et al. 2007). Here,
we neglect the stellar contribution to the X-ray background which is
negligible with respect to the AGN one (see also \citealt{Oh_01}), so that
$j(\nu,z)=j_{\rm BH}(\nu,z)$. The evolution of the X-ray background at
$\nu_0=150$~eV (dashed line) and 1~keV (dashed-dotted) is shown in
Figure~\ref{fig:back}. Even though a detailed comparison is beyond the scope
of this paper, we note that our results here are overall consistent with those
of \citet{RipamontiMapelliZaroubi_08} who used a set of semi-analitical models
to study X-ray heating from early black holes.

A strong upper limit to the AGN activity at high redshift is set by the
unresolved fraction of the observed cosmic X-ray background
\citep{SalvaterraHaardtVolonteri_07}. To make sure our mini-quasar population
does not conflict with this requirement, we compute the contribution of our
simulated BHs to the observed diffuse cosmic X-ray background in the 0.5-2 keV
and 2-8 keV bands. We find that this contribution is $\sim 0.9\times10^{-12}$
and $\sim 1.8\times10^{-12}$ erg s$^{-1}$ cm$^{-2}$ deg$^{-2}$, respectively,
that is $\sim 40$\% of the most recent estimates of the unresolved fraction
\citep{HickoxMarkevitch_07, Moretti_etal_09}. Even considering the expected
contribution of faint, $z<4$ AGNs missed by current deep surveys
\citep{VolonteriSalvaterraHaardt_06}, our calculation implies values well
below the observational limits.

\section{IGM heating}
\label{sec:heating}

Results from previous studies (e.g. \citealt{Madau_etal_04}) have shown 
that photons with energies $h\nu>150$~eV typically have a mean free path
larger than the average separation between sources; it is possible therefore
to divide the radiation field in a fluctuating component with
13.6~eV$<h\nu<150$~eV (which creates expanding patchy HII regions), and a
nearly uniform soft X-ray background component with 150~eV$<h\nu<2$~keV (which
ionizes the IGM homogeneously). Here we concentrate on the effect of the
latter component on the thermal evolution of the IGM.

We can calculate the energy rate per H atom (in units of [erg~s$^{-1}$]) at 
redshift $z$ as:
\begin{equation}
\dot{E}_x(z)=4\pi \int_{150 {\rm eV}}^{2 {\rm keV}} J(E,z)\sigma(E) dE,
\end{equation}
where $J(E,z)$, with $E=h \nu$, is computed by eq.~\ref{eq:back}.

Of the above energy, a fraction $f_{heat}$ goes into heating of the IGM and a
fraction $f_{ion}$ goes into H ionization
(\citealt{ShullvanSteenberg_85,ValdesFerrara_08})
\footnote{In \citet{ShullvanSteenberg_85}: $f_{heat}(z,x_{\rm HII})=0.9971
[1-(1-x_{\rm HII}^{0.2663})^{1.3163}]$ and $f_{ion}(z,x_{\rm HII})=0.3908
(1-x_{\rm HII}^{0.4092})^{1.7592}$.  In \citet{ValdesFerrara_08}:
$f_{heat}(z,x_{\rm HII})=1 - 0.8751 (1 - x_{\rm HII}^{0.4052})$ and
$f_{ion}(z,x_{\rm HII})=0.3846 (1 - x_{\rm HII}^{0.5420})^{1.1952}$.}.  We
have verified that by using the two parameterizations of $f_{heat}$ the overall 
discrepancy in our results (in terms of IGM temperature) is at most a few percent.
In the following, we adopt the results by \cite{ValdesFerrara_08}.

The evolution of the IGM temperature, $T_{IGM}$, is followed as in
\cite{CiardiSalvaterra_07} (based on \citealt{ChuzhoyShapiro_07}),
i.e. $T_{IGM}$ is calculated with {\tt RECFAST} \citep{SeagerSasselovScott_99}
until the first sources of radiation turn on.
Subsequentely, taking into account the presence of the heating from
X-ray photons and neglecting the contribution from deuterium, which turned out
to be negligible, the evolution of $T_{IGM}$ is regulated by:

\begin{equation}
\frac{dT_{IGM}}{dt}=\frac{2}{3k} \left(H_\alpha+H_x \right) - \frac{4T_{IGM}}{3t}.
\label{eq:temp}
\end{equation}
Here $H_\alpha$ and $H_x=f_{heat}(z,x_{\rm HII}) \dot{E}_x(z)$ are the heating
rates per H atom due to Ly$\alpha$ and X-ray photons, respectively.  We refer
the reader to the original paper (\citealt{CiardiSalvaterra_07}) for the
derivation of $H_\alpha$.  $\dot{N}_\alpha$ (the number of photons that pass
through the Ly$\alpha$ resonance per H atom per unit time) necessary to
calculate $H_\alpha$ has been derived from the model described in the previous
Section.

It should be noted that the impact of Ly$\alpha$ photon scattering on the evolution
of the IGM temperature has been recently revised by e.g. \cite{ChenMiralda-Escude_04},
\cite{Hirata_06}, \cite{PritchardFurlanetto_06}
and \cite{ChuzhoyShapiro_07}. \cite{ChenMiralda-Escude_04} included atomic thermal
motion in their calculations, finding a heating rate several orders of magnitude lower 
than the previous estimate by \cite{MadauMeiksinRees_97}. In addition, while ``continuum''
photons (with frequency between the Ly$\alpha$ and Ly$\beta$) heat the gas, ``injected'' 
photons (which cascade into the Ly$\alpha$ from higher atomic resonances) cool the gas, 
resulting in an effective cooling from Ly$\alpha$ photons at temperatures above 10~K. 
 \cite{ChuzhoyShapiro_07} though, showed that the cascade which follows absorption of 
photons in resonances higher than the Ly$\alpha$ happens via the 2s level rather than the
2p level. Thus, the number of ``injected'' photons and their cooling efficiency are reduced 
compared to the estimate of \cite{ChenMiralda-Escude_04} and Ly$\alpha$ photon scattering
can be an efficient heating source also at temperatures higher than 10~K. 
Here we follow the calculation of \cite{ChuzhoyShapiro_07}.

As $H_x$ is also a function of the ionization fraction, eq.~\ref{eq:temp}
needs to be solved together with the following equation:
\begin{equation}
\frac{dx_{\rm HII}}{dt}=f_{ion}(t,x_{\rm HII}) \dot{N}_x \frac{n_{\rm H}}{n_{\rm HI}} + k_{coll} n_e x_{\rm HI} -
k_{rec} n_e x_{\rm HII},
\label{eq:ion}
\end{equation}
where $x_{\rm HII}$ and $x_{\rm HI}$ are the HII and HI fraction respectively,
$n_{\rm H}$ and $n_{\rm HI}$ are the number density of H and HI respectively,
$k_{coll}$ and $k_{rec}$ are the collisional and recombination rates,
$\dot{N}_x$ is the rate of X-ray photons per H atom.

\begin{figure}
\centering
\includegraphics[width=0.5\textwidth]{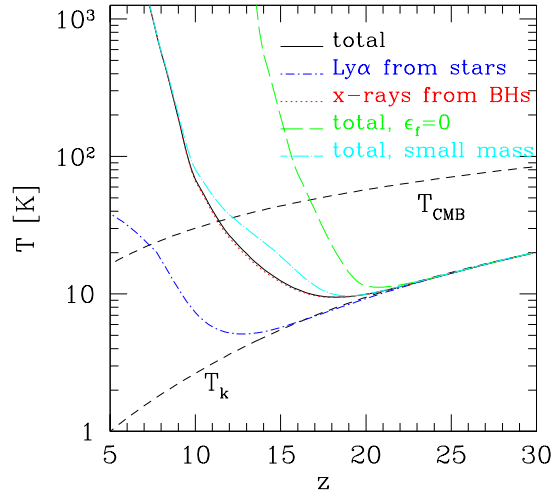}
\caption{Temperature evolution as a function of redshift: total (solid black
line), total in the absence of black holes feedback (long-dashed green line),
total with an extension to lower masses of the contribution of stellar type
sources (long-dashed-dotted cyan line) as determined by X-ray heating from
black holes only (dotted red line) and by Ly$\alpha$ photons from stars only
(short-dashed-dotted blue line). The upper (lower) dashed curve is the CMB
temperature (the IGM temperature in the absence of any heating mechanisms).}
\label{fig:temp_igm}
\end{figure}

In Figure~\ref{fig:temp_igm} the evolution of the IGM temperature as given by
eqs.~\ref{eq:temp} and~\ref{eq:ion} is shown, together with the CMB
temperature, $T_{\rm CMB}$, and the IGM temperature, $T_k$, in the absence of
any heating mechanisms. The solid curve includes the contribution to the
heating of X-rays from BHs, Ly$\alpha$ photons from both BHs and stars. It is
clear from the Figure that the contribution from the Ly$\alpha$ photons is
irrelevant for the evolution of the IGM temperature, which is instead
dominated by the X-ray heating (see Sec~\ref{sec:summary} for a more extensive
discussion). We note that the contribution to the Ly$\alpha$ photons from the
BHs is negligible compared to that from the stars.

\section{Spin and differential brightness temperature}
\label{sec:21cm}

As the physics behind the emission/absorption of the 21~cm line has been
discussed extensively by several authors (for a review see
e.g. \citealt{FurlanettoOhBriggs_06}), here we just write the relevant
equations, following \cite{CiardiSalvaterra_07}.  The evolution of the spin
temperature, $T_s$, can be written as \cite{ChuzhoyShapiro_06}:
\begin{equation}
T_s=\frac{T_{\rm CMB} + (y_{\alpha,eff} + y_c) T_{\rm
IGM}}{1+y_{\alpha,eff}+y_c},
\end{equation}
where $T_{\rm CMB}$ is the CMB temperature;
\begin{equation}
y_c=\frac{T_\star}{A_{10}T_{\rm IGM}}(C_{\rm H} + C_e +C_p),
\end{equation}
is the coupling efficiency due to collisions with H atoms, electrons and
protons, $T_\star=0.068$~K is the temperature corresponding to the energy
difference between the singlet and triplet hyperfine levels of the ground
state of neutral hydrogen and $A_{10}=2.85 \times 10^{-15}$~s$^{-1}$ is the
spontaneous emission rate. For the de-excitation rates $C_{\rm H}$, $C_e$ and
$C_p$ we have adopted the fits used by Kuhlen, Madau \& Montgomery (2006; see
also the original papers by \citealt{Smith_66,
AllisonDalgarno_69,Liszt_01,Zygelman_05}), i.e. $C_{\rm H}= n_{\rm H} \kappa$,
$C_e=n_e \gamma_e$ and $C_p=3.2 n_p \kappa$ with $n_{\rm H}$, $n_e$, $n_p$
hydrogen, electron and proton number density, $\kappa=3.1 \times 10^{-11}
T_{\rm IGM}^{0.357} {\rm exp}(-32/T_{\rm IGM})$~cm$^3$~s$^{-1}$ and ${\rm
log}(\gamma_e/1 {\rm cm}^3{\rm s}^{-1})= -9.607+0.5{\rm log}(T_{\rm IGM}) {\rm
exp}[-({\rm log}T_{\rm IGM})^{4.5}/1800]$. Finally,
\begin{equation}
y_{\alpha,eff}=y_\alpha {\rm e}^{-0.3(1+z)^{1/2} T_{\rm IGM}^{-2/3}} \left( 1+\frac{0.4}{T_{\rm IGM}}\right)^{-1},
\end{equation}
is the effective coupling efficiency due to Ly$\alpha$ scattering which takes
into account the back-reaction of the resonance on the Ly$\alpha$ spectrum and
the effect of resonant photons other than Ly$\alpha$, and
$y_\alpha=P_{10}T_\star/(A_{10} T_{\rm IGM})$, with $P_{10}\sim 10^9
J_\alpha$~s$^{-1}$ radiative de-excitation rate due to Ly$\alpha$ photons.

\begin{figure}
\centering
\includegraphics[width=0.5\textwidth]{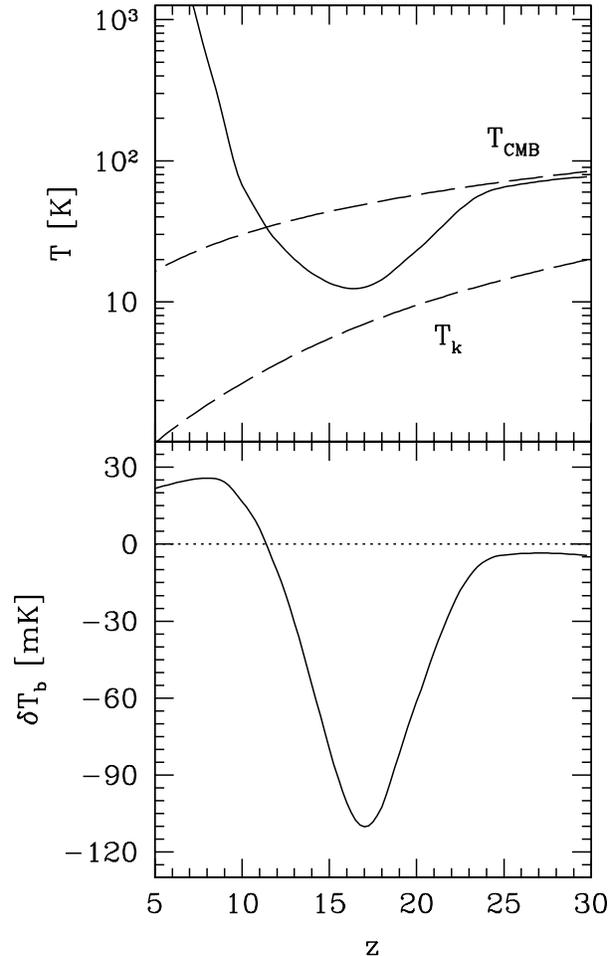}
\caption{{\it Upper panel}: Evolution of the spin temperature (solid
line). $T_{\rm CMB}$ and the IGM temperature in the absence of heating
mechanisms, $T_k$, are plotted as upper and lower dashed lines,
respectively. {\it Lower panel}: Evolution of the differential brightness
temperature. Lines are the same as in the upper panel.  }
\label{fig:temp_21cm}
\end{figure}

In the upper panel of Figure~\ref{fig:temp_21cm} the spin temperature is shown
as a function of redshift together with $T_{\rm CMB}$ and the IGM temperature
in the absence of heating mechanisms $T_k$ (upper and lower dashed lines,
respectively).  As the intensity required for a Ly$\alpha$ background to be
effective in decoupling the spin temperature from $T_{\rm CMB}$ is $J_\alpha
\simgt 10^{-22} (1+z)$~ergs~cm$^{-2}$~s$^{-1}$~Hz$^{-1}$~sr$^{-1}$
(e.g. \citealt{CiardiMadau_03}, see Fig.~\ref{fig:back}), $T_s$ gets coupled
to $T_{\rm IGM}$ as early as $z \sim 19$, but only at $z \sim 11$ does it
become larger than $T_{\rm CMB}$ due to X-ray heating.

The differential brightness temperature, $\delta T_b$, between the CMB and a
patch of neutral hydrogen with optical depth $\tau$ and spin temperature $T_s$
at redshift $z$, can be written as:
\begin{equation}
\delta T_b=(1-{\rm e}^{-\tau}) \frac{T_s-T_{\rm CMB}}{1+z}.
\end{equation}
For a gas at the mean IGM density, the optical depth is:
\begin{equation}
\tau(z)=\frac{3A_{10}\lambda^3}{32 \pi} \frac{T_\star}{T_s} \frac{n_{\rm HI}(z)}{H(z)},
\end{equation}
where $\lambda$ is the wavelength of the transition, $n_{\rm HI}$ is the
number density of neutral hydrogen and $H(z)$ is the Hubble constant.

In the lower panel of Figure~\ref{fig:temp_21cm} $\delta T_b$ is plotted for
the spin temperature calculated above, reflecting its behavior, with a signal
in emission (absorption) expected for $z \simlt (\simgt) 11$.

\section{Discussion and summary}
\label{sec:summary}

In this paper we have estimated the effect of X-ray and Ly$\alpha$ photons on
the IGM temperature. This was done calculating the photon production from a
population of both stars and quasars based on the self-consistent modeling of
these sources and associated feedback processes in the simulations by
\citet{PelupessyDiMatteoCiardi_07}. 
Such photon production has then been used to determine the evolution of the
IGM temperature with a semi-analytic approach.
The calculations assume a gas at the mean
density whose ionization history is regulated by the effect of the X-ray
photons. Thus, they are strictly valid as long as the filling factor of gas
ionized by UV photons is small or in regions of the IGM that have not been
reached by UV-ionizing radiation.  For reference, the filling factor of
ionized regions, under the assumption that it is independent from the effect
of the X-rays, assuming a gas clumping factor of 10 and an escape fraction of
ionizing photons from stellar type sources of 10\% is $\sim 0.004, 0.07, 1$ at
$z=15, 10, 7.1$. It should be noted here that the bulk of the UV photons is
produced by stars rather than BHs, while the contribution to the ionization
fraction from X-rays is only a few percents.

The results presented of course depend quite strongly on the parameters and
assumptions adopted in the simulations described in
Section~\ref{sec:simul}. As already mentioned, while e.g. the value of the
spin parameter has a negligible effect on the final outcome, the presence of
AGN feedback (but not the specific value of $\epsilon_f$, as long as $\ne 0$)
is crucial. In our reference run we have used a vale $\epsilon_f=0.5$, which
corresponds to coupling only $5\%$ of the radiated energy. In
\citet{PelupessyDiMatteoCiardi_07} we have shown that no significant
difference in the BH accretion rates are seen for $\epsilon_f=0.5$, but that
of course cases with $\epsilon_f=0$ can imply much larger accretion rates (and
associated X-ray photons).  For the sake of illustration, we have performed
the same calculations also for the extreme case without feedback, i.e. for
$\epsilon_f=0$. Note that for our calculation this is a completely 
unrealistic model, which we consider only as a parametric study.
The result is shown in Figure~\ref{fig:temp_igm} as
long-dashed line. As expected, the value of the IGM temperature raises much
more quickly than in the reference case and becomes larger than $T_{\rm CMB}$
already at $z \sim 16$. This would result also in an earlier 21~cm signal in
emission.  This case is both unrealistic and also results in a contribution to
the unresolved X-ray background which is orders of magnitudes larger than the
available observational limits and thus can be discarded.

So it remains that our results depend most strongly on the choice of
$M_{h,min}$ and $M_{h,max}$ in eqs.~\ref{eq:jBH} and~\ref{eq:sfr}, which in
our case we have taken from the masses of the simulated halos. While larger
values of $M_{h,max}$ are expected to have a negligible impact because of the
paucity of such objects at the redshifts of interest, a smaller $M_{h,min}$
(and its associated high abundance) would result in higher photon
production. While extending eq.~\ref{eq:sfr} to smaller masses is
straightforward, it may not be appropriate as the contribution from these low
mass halos depends on more complicated forms and strength of the feedback
active in the relevant range of redshift (for a review on feedback see
\citealt{CiardiFerrara_05}). Although a consensus on the detailed effects of
feedback (in particular radiative feedback) on small structure formation has
not been reached yet, most authors agree that the presence of feeback delayes
or suppresses the formation of small mass objects, and thus associated star
formation, depending on a variety of physical quantities such as redshift,
halo mass and molecular hydrogen content (see e.g.
\citealt{MachacekBryanAbel_01,SusaUmemura_06,OSheaNorman_07,Whalen_etal_08a}).
A proper treatment of such feedback though is beyond the scope of the present
paper.  Similarly the emission properties of mini-QSOs hosted in such small
halos is largely unknown and may not be described by the average AGN spectrum
here adopted. Thus, we have neglected the contribution of these objects to the
total X-ray photon production (see \citealt{Madau_etal_04, Dijkstra_etal_04,
Salvaterra_etal_05} for a discussion about the effect of mini-QSOs in the
early universe).

As a reference though, an upper limit for the contribution of Ly$\alpha$
photon to the IGM temperature is plotted in Figure~\ref{fig:temp_igm}. Here
(long-dashed-dotted cyan line) we show the effects accounting for the
contribution to the star formation rate of objects with mass as small as
$10^7$~M$_\odot$. As expected, in this case, the IGM temperature starts to
raise at an earlier time due to Ly$\alpha$ heating and becomes larger than
$T_{\rm CMB}$ at $z \sim 12$, while X-ray photons dominate heating at $z
\simlt 10$. Note however, this considers that these halos will not produce any
X-ray photons, i.e. the accretion onto their black holes must be completely
quenched. This is likely a conflicting requirement, as both star formation and
black hole accretion will somewhat depend on the same gas supply. To summarize
this point, in order to have 21~cm line in emission, X-ray heating from BHs
and/or extremely efficient star formation (albeit, as discussed above,
somewhat unfeasible) in halos with $M_{h} < 10^8 \Msun$ is required.  The
heating due to stars only in larger halos as shown in
Figure~\ref{fig:temp_igm}, is however unable to bring the gas temperature
above $T_{\rm CMB}$ before the reionization is completed.

To summarize the main results of this work, we find that:
\begin{itemize}
\item unless the Ly$\alpha$ photon contribution from stars in small mass
objects is somehow strongly enhanced in comparison to the X-ray contribution
from BHs in the same objects, the IGM heating is always dominated by X-rays;

\item in this case, the transition redshift between a 21~cm signal in
absorption and in emission could be used to constraint BHs accretion
properties;
\item for a case consistent with the available limits on the unresolved
fraction of the X-ray background in the 0.5-2~keV and 2-8~keV bands, the IGM
temperature becomes larger than the CMB temperature (and thus an associated
21~cm signal should be expected in emission) at $z \simlt 11.5$.
\end{itemize}

\bibliographystyle{mn2e}
\bibliography{heating.bib}

\label{lastpage}

\end{document}